\documentclass[aps,prb,twocolumn]{revtex4}
\usepackage{epsf}
\usepackage[intlimits]{amsmath}
\usepackage{amssymb}
\usepackage{graphicx}

\newcommand{\bacu}{BaCuSi$_2$O$_6$}

\begin{document}

\title {
Quantum phase transitions and dimensional reduction \\
in antiferromagnets with inter-layer frustration
}

\author{Oliver R\"osch}
\author{Matthias Vojta}
\affiliation{Institut f\"ur Theoretische Physik,
Universit\"at zu K\"oln, Z\"ulpicher Str. 77, 50937 K\"oln, Germany}
\date{September 3, 2007}

\begin{abstract}
For magnets with a fully frustrated inter-layer interaction,
we argue that the quantum phase transitions from a paramagnetic to an antiferromagnetic
ground state, driven by pressure or magnetic field,
are asymptotically three-dimensional,
due to interaction-generated non-frustrated inter-layer couplings.
However, the relevant crossover scale is tiny,
such that two-dimensional behavior occurs in an experimentally relevant
low-temperature regime.
In the pressure-driven case the phase transition may split,
in which case an Ising symmetry related to inter-layer bond order is broken
before magnetism occurs.
We discuss the relation of our results to recent experiments on \bacu.
\end{abstract}

\maketitle

%%%%%%%%%%%%%%%%%%%%%%%%%%%%%%%%%%%%%%%%%%%%%%%%%%%%%%%%%%%%%%%%%%%%%%%

In modern condensed matter physics, reduced dimensionality presents a
fascinating avenue to novel effects arising from strong fluctuations.
The standard realization of reduced dimensionality is via systems of
chains or planes with a weak three-dimensional (3d) coupling in suitably
structured materials.
In such a situation, the physics is one-dimensional (1d) or two-dimensional (2d)
at elevated energies (or temperatures), while it becomes 3d
in the low-energy limit.

Recent experiments near magnetic quantum critical points (QCP) -- both
in heavy-fermion metals\cite{stockert,ybrhsi,rmp}
and insulating dimer magnets\cite{sebastian} --
have raised speculations about a rather different route to reduced dimensionality,
namely through geometric frustration.
The idea is that fully frustrated 3d interactions (being not necessarily weak)
between 2d units effectively vanish in a well-defined low-energy limit,
which in particular can be realized at a QCP.
Then, it has been suggested that true 2d behavior
could be observed at lowest energies.\cite{sebastian}
%, which may cross over to 3d behavior at intermediate energies.

Particularly interesting are the results on Mott-insulating quantum paramagnets,
where the spin gap can be closed either by application of a magnetic field or
pressure.\cite{rueggpressure,rueggtlcucl,bacusio,sebastian}
The field-driven quantum phase transition (QPT),
from a paramagnet to an XY-ordered antiferromagnet (AF) at a field $H=H_{c1}$,
belongs to the universality class of the dilute Bose gas, and the
finite-temperature transition can be understood as Bose-Einstein
condensation of magnons.
In \bacu, the transition line has been found\cite{sebastian}
to follow $T_c \propto (H-H_{c1})^\psi$, with a shift exponent $\psi=1$ characteristic of
a 2d QCP, and indications for a crossover
to the 3d value $\psi=2/3$ at higher $T$.
As \bacu\ has a body-centered tetragonal (bct) structure of Cu dimers,
with a frustrated inter-layer coupling,
the results have been interpreted as dimensional reduction at a QCP arising
from geometric frustration.

On the theory side, an investigation\cite{coleman} of the ordered state
of a frustrated double-layer Heisenberg model using spin-wave theory
concluded that inter-layer order is stabilized
through an order-from-disorder mechanism, thus rendering dimensional reduction
ineffective.
In fact, very similar results were obtained much earlier in spin-wave studies of
frustrated bct magnets.\cite{rastelli,shender}
However, the situation is different at a QCP,\cite{sebastian,batista}
as the fluctuation-generated inter-layer coupling can be expected to be
proportional to the square of the order parameter and hence absent at the QCP,
supporting the interpretation of dimensional reduction from frustration.
However, as we show below, higher-order processes turn out to be relevant
in determining the nature of the phase transitions.

\begin{figure}[b]
\epsfxsize=3.2in
\centerline{\epsffile{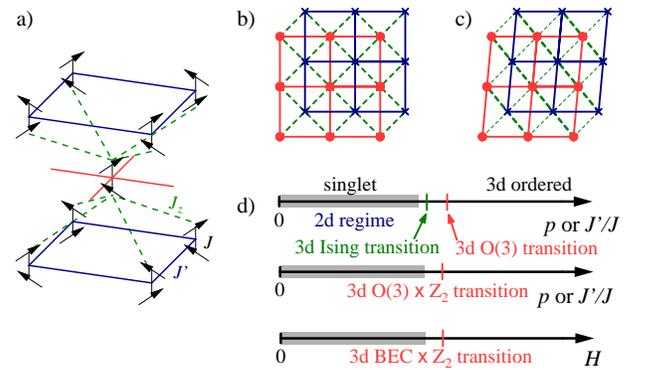}}
\caption{
(Color online)
a) bct lattice of dimers.
b) top view: two planes with sites shown as circles/crosses,
and in-plane (inter-plane) coupling as solid (dashed) lines.
c) Illustration of the Z$_2$ symmetry breaking,
where the diagonal inter-plane bonds develop a spontaneous asymmetry (see text),
together with the simplest lattice compatible distortion.
d) $T=0$ phase diagrams for the coupled-dimer model (\ref{H}),
for the pressure-driven case with and without splitting (see text)
and the field-driven case.
}
\label{fig:schem1}
\end{figure}

The purpose of this paper is an investigation of QPT
in such Mott-insulating magnets with inter-layer frustration,
using a detailed symmetry analysis, field theory and bond-operator approaches.
Our main results can be summarized as follows:
(i) In general, geometric frustration {\em cannot} lead to dimensional reduction
at asymptotically low energies, due to fluctuation-generated unfrustrated interactions.
But their energy scale is tiny at quantum criticality, rendering
2d behavior observable even at very low temperatures.
(ii) For the bct lattice, the ordered state breaks both the magnetic and an Ising
symmetry. In the pressure-driven case, two scenarios are possible.
Either there is a single transition breaking both symmetries,
or the transitions is split:
the Ising symmetry is broken first, which relieves the inter-layer frustration,
and the subsequent magnetic transition is conventional,
Fig.~\ref{fig:schem1}.

%%%%%%%%%%%%%%%%%%%%%%%%%%%%%%%%%%%%%%%%%%%%%%%%%%%%%%%%%%%%%%%%%%%%%%%

{\it Model Hamiltonian.}
To be specific, we will consider a coupled-dimer system on a bct lattice
(but we believe our results to be more general), with the Hamiltonian
\begin{eqnarray}
\label{H}
{\cal H} &=&
  J                          \sum_{in} {\vec S}_{in1} \!\cdot\! {\vec S}_{in2}
+ J' \! \sum_{\langle ij\rangle nm} \! {\vec S}_{inm} \!\cdot\! {\vec S}_{jnm} \\
&+&
  \!\!\sum_{i\Delta n m m'} \!\!\! J_z^{mm'}{\vec S}_{inm} \!\cdot\! {\vec S}_{i+\Delta,n+1,m'}
- {\vec H}\!\cdot\! \sum_{inm} {\vec S}_{inm}
\nonumber
\end{eqnarray}
where $m=1,2$ labels the two spins 1/2 of each dimer, $i,j$ are the dimer site indices in each
layer, and $n$ is the layer index.
$J$ and $J'$ are the AF intra-dimer and in-plane inter-dimer couplings,
respectively, while $J_z^{mm'}$ represent the frustrated inter-layer couplings
(with some specific inter-dimer structure given by the $mm'$ dependence).
The $\sum_\Delta$ runs over four sites such that the sites $(in)$ and $(i+\Delta,n+1)$
are nearest neighbors in $z$ direction.
This Hamiltonian is assumed to be relevant for the material \bacu\cite{bacusio}
(neglecting here the orthorhombic distortions in the low-temperature
phase\cite{samulon}).

%%%%%%%%%%%%%%%%%%%%%%%%%%%%%%%%%%%%%%%%%%%%%%%%%%%%%%%%%%%%%%%%%%%%%%%

{\it Phases.}
For $J\gg J',|J_z|$, the zero-field ground state of $\cal H$ is a paramagnetic singlet,
with gapped triplet excitations.
If $J'$ dominates, an AF phase with in-plane ordering
wavevector ${\vec Q}=(\pi,\pi)$ is established, and the order in $z$ direction is frustrated,
as discussed below.
(For large $|J_z|$ the in-plane order is ferromagnetic.)
Applying a field to the large-$J$ quantum paramagnet leads to a Zeeman
splitting of the triplet excitations.
At a critical field $H_{c1}$, the gap of the lowest mode closes, and
a QPT to a gapless canted phase occurs.
Upon further increasing the field, the system is driven into
a fully polarized state at $H_{c2}$.
The phase diagram at $T=0$ is thus similar to that
of the much-studied bilayer Heisenberg model.\cite{sandvik,kotov}

%%%%%%%%%%%%%%%%%%%%%%%%%%%%%%%%%%%%%%%%%%%%%%%%%%%%%%%%%%%%%%%%%%%%%%%

{\it AF phase: Order from disorder.}
While AF ordered planes of classical moments on the bct lattice are
uncoupled, zero-point fluctuations of quantum spins lift this large degeneracy.
A spin-wave calculation, for spins $S$ with couplings $J'$, $J_z$
and a helical order with wavevector $(\pi,\pi,Q_z)$,
yields an inter-layer contribution to the ground-state energy
of the form $(-J_z^2 S/J' \cos^2 Q_z)$,
favoring {\em collinear} (i.e. parallel {\em or} antiparallel) order between adjacent
planes.\cite{rastelli,coleman}
In addition, higher-order terms, in a calculation for a more general ordering pattern,
actually stabilize AF order between {\em 2nd}-neighbor planes.\cite{shender}
% the energy contributions are of order $J_z^4/(JS)^3$ and $J_z^6/(J^5 S)$.
%
Thus, fully 3d order is stabilized within the stacks of ``even'' and ``odd'' planes,
but a residual Z$_2$ degeneracy is left intact, corresponding to a spin inversion
in every second plane --
this represents a true symmetry of the AF on the bct lattice,
and will be spontaneously broken in the ordered phases.

%%%%%%%%%%%%%%%%%%%%%%%%%%%%%%%%%%%%%%%%%%%%%%%%%%%%%%%%%%%%%%%%%%%%%%%

{\it Symmetries and magnetic order parameter.}
%We proceed by investigating low-energy antiferromagnetic fluctuations
%within an order-parameter language.
The in-plane magnetism is unfrustrated, allowing to define an
order parameter ${\vec \phi}_n(\vec r_\parallel)$, where $\vec r_\parallel$ is the in-plane coordinate and $n$ the
layer index, with the local magnetization operator given by
${\vec m}_n({\vec r_\parallel}) = \exp(i {\vec Q \cdot \vec r_\parallel}) {\vec \phi}_n(\vec r_\parallel)$.
Expanding the tight-binding dispersion on the bct lattice around $\vec Q$,
i.e. taking the in-plane continuum limit,
we arrive at a $\phi^4$ theory for the magnetic fluctuations:
\begin{eqnarray}
{\cal S}_\phi &=&
\int d\tau d^2 k_\parallel \sum_n \Big[
( m_\phi + c^2 k_\parallel^2) {\vec \phi}_n^2({\vec k_\parallel})  \nonumber\\
&&+ \eta c^2 k_x k_y {\vec \phi}_n \cdot {\vec \phi}_{n+1}
\Big]
+ {\cal S}_{\phi 4} + {\cal S}_{\phi \rm dyn}
\label{phi4}
\end{eqnarray}
where ${\vec \phi}_n({\vec k_\parallel})$ is the real order-parameter field,
% after Fourier transformation w.r.t.\ the in-plane coordinates,
$\vec k_\parallel=0$ now corresponds to physical in-plane momentum $(\pi,\pi)$,
$c$ is a velocity,
and $\eta$ represents the spatial anisotropy (i.e. $\eta \sim J_z/J'$).
Further, ${\cal S}_{\phi \rm dyn}$ encodes the dynamics of the spin fluctuations
(which depends on whether an external field is present or absent),
and ${\cal S}_{\phi 4}$ is the local quartic self-interaction,
%\begin{equation}
$
{\cal S}_{\phi 4} = u_0 \int d\tau d^2 r_\parallel \sum_n
[{\vec \phi}_n^2(\vec r_\parallel)]^2.
$
%\end{equation}
Effects of the geometric frustration completely dominate the inter-layer dispersion:
it is not only suppressed at zero wavevector,
but also acquires a sign alternating from quadrant to quadrant away from it.\cite{sebastian}

The action (\ref{phi4}) represents the ``bare'' order-parameter dynamics;
additional symmetry-allowed terms will be generated
upon integrating out high-energy modes.
We analyze the symmetries of the problem:
In the zero-field paramagnetic phase,
time and space inversion as well as SU(2) spin rotations are unbroken.
The system is also invariant under 90-degree in-plane rotations,
however, the frustrated geometry dictates that this is accompanied by a
relative sign change of the order parameter in two neighboring planes:
\begin{equation}
k_x \rightarrow k_y\,,~
k_y \rightarrow -k_x\,,~
{\vec \phi}_n \rightarrow (-1)^n {\vec \phi}_n \,.
\label{sym}
\end{equation}
Obviously, the action ${\cal S}_\phi$ (\ref{phi4}) is invariant under this transformation.
Now we can list the lowest-order additional terms allowed and forbidden by symmetry:
\begin{eqnarray}
{\rm Allowed\!:}&&\!\!\!
\eta' \vec\phi_n \!\cdot\! \vec\phi_{n+2},~
u_1 (\vec\phi_n \!\cdot\! \vec\phi_{n+1})^2,~
u_2 \vec\phi_n^2 \vec\phi_{n+1}^2,
..
\label{allowed} \\ \!\!\!
{\rm Forbidden\!:}&&\!\!
\vec\phi_n \cdot \vec\phi_{n+1},~
k_x \vec\phi_n \cdot \vec\phi_{n+1},
..
\label{forbidden}
\end{eqnarray}
Thus, single-particle hopping between neighboring planes
must be suppressed at least by a factor $k_x k_y$ (i.e. it is frustrated),
as in (\ref{phi4}).
However, single-particle hopping between 2nd-neighbor planes is
allowed even at ${\vec k_\parallel}=0$,
as are density interactions between $\phi_n$ and $\phi_{n+1}$.

\begin{figure}[t]
\epsfxsize=3in
\centerline{\epsffile{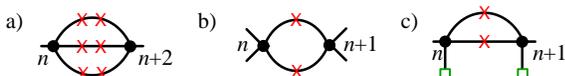}}
\caption{
(Color online)
Diagrams relevant for the order-parameter theory ${\cal S}_\phi$.
Lines are $\phi_n(\vec k_\parallel)$ propagators,
the circle is the local four-point vertex ($\propto u_0$),
the cross is the interlayer hopping ($\propto \eta$).
a) Unfrustrated vertical 2nd-neighbor coupling --
this is the $\eta'$ term in (\ref{allowed}) which is responsible for
3d behavior at lowest energies.
b) Inter-layer density interaction, generating the $u_1$ and $u_2$ terms
in (\ref{allowed}). $u_1<0$ leads to collinear inter-plane
spin correlations.
c) Vertical near-neighbor coupling in the ordered phase
(the open square denotes the condensate).
}
\label{fig:dgr1}
\end{figure}

The first two terms in (\ref{allowed}) are of
crucial importance for the following analysis;
they obviously correspond to additional terms to the
microscopic Hamiltonian (\ref{H})
\begin{eqnarray}
{\cal H}_{\rm zz} &=& \sum_{inmm'} \! J_{\rm zz}^{mm'}
{\vec S}_{inm} \!\cdot\! {\vec S}_{i,n+2,m'},
\label{hzz} \\
{\cal H}_{\rm coll} &=&
  \sum_{i\Delta n m m'} \!\! J_{\rm coll}^{mm'}
  ({\vec S}_{inm} \!\cdot\! {\vec S}_{i+\Delta,n+1,m'})^2,
\label{hcoll}
\end{eqnarray}
which can be present in any real material.
However, even if absent in the ``bare'' model, they will be perturbatively generated.
The corresponding lowest-order diagrams, quadratic in the interaction vertex $u_0$,
are shown in Fig.~\ref{fig:dgr1}a,b.
Even if the $u$ terms are (marginally) irrelevant in the RG sense at the QCP
of the theory (\ref{phi4}) (which is the case at the $H_{c1}$ critical
point), the $\eta' \vec\phi_n \cdot \vec\phi_{n+2}$ term is strongly relevant.

Thus we arrive at a central result:
There exists an energy scale in the $\phi$ dispersion, $E_{z} \sim \eta'$,
corresponding to an {\em unfrustrated} 3d coupling in the paramagnetic phase,
which is finite also at the QCP.
Then, the behavior of observables at energies or temperatures below $E_{z}$ will be
fully 3d (with a dimensional crossover to 2d behavior above $E_{z}$),
and {\em no} true dimensional reduction occurs.

A straightforward calculation of the diagram in Fig.~\ref{fig:dgr1}b
gives a negative coefficient $u_1 \propto -\eta^2$ for the $(\vec\phi_n \cdot \vec\phi_{n+1})^2$
interaction, thus stabilizing collinear spin correlations between adjacent planes,
consistent with the spin-wave result for the ordered phase.
In fact, an effective Hamiltonian of the form (\ref{hcoll}),
with $J_{\rm coll} \propto - J_z^2/(J' S^3)$,
was proposed before in\cite{coleman} to mimic the collinear order-from-disorder
mechanism within linear spin-wave theory.

%%%%%%%%%%%%%%%%%%%%%%%%%%%%%%%%%%%%%%%%%%%%%%%%%%%%%%%%%%%%%%%%%%%%%%%

{\it Ising order parameter.}
As emphasized above, the magnetically ordered phases break an additional Z$_2$
symmetry -- in the $\phi$ language this degree of freedom
corresponds to the sign of $\langle\vec\phi_n\rangle\cdot\langle\vec\phi_{n+1} \rangle$.
It is useful to introduce an associated (inter-plane) Ising order parameter
$\Psi_{n+1/2}({\vec r}_\parallel)$.
It transforms as a singlet under spin rotations, hence symmetry dictates that
$\Psi$ can be described by a standard $\Psi^4$ theory ${\cal S}_\Psi$,
similar to ${\cal S}_\phi$ (\ref{phi4}), but with mass $m_\Psi$ and
an {\em unfrustrated} vertical hopping $\Psi_{n-1/2} \Psi_{n+1/2}$.
%\begin{eqnarray}
%{\cal S}_\psi &=&
%\int d\tau d^2 k_\parallel \Big[
%\sum_n ( m_\Psi + k_\parallel^2) \Psi_{n+1/2}^2({\vec k_\parallel})  \nonumber\\
%&&+ \eta_\Psi \sum_{n} \Psi_{n-1/2} \Psi_{n+1/2}
%\Big]
%+ {\cal S}_{\Psi 4} + {\cal S}_{\Psi \rm dyn} \,.
%\label{psi4}
%\end{eqnarray}
The physical content of $\Psi$ becomes clear from its interaction with $\vec\phi$,
which is of Yukawa type:
\begin{eqnarray}
{\cal S}_{\phi\Psi} &=& \lambda
\int d\tau d^2 k_\parallel
\sum_n  \Psi_{n+1/2} \, \vec\phi_n \cdot \vec\phi_{n+1}
\label{phipsi}
\end{eqnarray}
with $\lambda$ a coupling constant.
(Additional couplings $\Psi \vec\phi_n^2$
do not modify the physics to be discussed below.)

%%%%%%%%%%%%%%%%%%%%%%%%%%%%%%%%%%%%%%%%%%%%%%%%%%%%%%%%%%%%%%%%%%%%%%%

{\it One vs. two transitions.}
The full theory ${\cal S}_\phi+{\cal S}_\Psi+{\cal S}_{\phi\Psi}$
admits two distinct scenarios:
(A) A single transition driven by the condensation of $\phi$ --
here, the coupling $\lambda$ generates a non-zero expectation value for $\Psi$ as well,
because the $u_1$ term in ${\cal S}_\phi$ leads to non-zero
$\langle \vec\phi_n\cdot\vec\phi_{n+1} \rangle$ (Fig.~\ref{fig:dgr1}c).
(B) Two transitions: First, $\Psi$ condenses, which only modifies the quadratic part
of the $\phi$ action, and $\phi$ orders in a second, subsequent transition.

Scenario (B) is appealing:
From (\ref{phipsi}) it can be seen that a $\Psi$ condensate generates
an {\em unfrustrated} vertical hopping of $\phi$
through the term $\lambda\langle\Psi\rangle\vec\phi_n \cdot \vec\phi_{n+1}$.
Thus, the $\Psi$ ordering transition removes the magnetic inter-plane
frustration.
In the lattice model, this can be understood as spontaneous bond order
which modulates the vertical magnetic couplings within each unit cell
and can easily couple to lattice distortions,
Fig.~\ref{fig:schem1}c, i.e.,
$\Psi$ ordering is a structural phase transition.

In zero field, a model with negative $u_1$ may follow scenario (B)
for the pressure-driven transition:
$\Psi$ can be understood as composed of two $\phi$ quanta, hence $m_\Psi \sim 2 m_\phi$.
The $u_1$ term mediates an attraction between $\phi_n$ and $\phi_{n+1}$.
Approaching the $\phi$ ordering transition,
$m_\Psi$ can become smaller than $m_\phi$, % the {\em single}-particle gap,
implying that $\Psi$ condenses before $\phi$.
This requires a sufficiently strong $|u_1|$
(otherwise no true two-particle bound state is generated,
and the transition remains in scenario (A)).

%%%%%%%%%%%%%%%%%%%%%%%%%%%%%%%%%%%%%%%%%%%%%%%%%%%%%%%%%%%%%%%%%%%%%%%

{\it Lattice theory.}
We have studied the coupled-dimer model (\ref{H}) using the bond-operator
approach.\cite{bondop}
Starting from a singlet product state on dimer sites $i$,
$\prod_i |i,s\rangle$,
we define bosonic operators $t^\dagger_{i\alpha}$ which create local
triplet excitations $|i,\alpha\rangle = t^\dagger_{i\alpha}\left|i,s\rangle\right.$
where $\alpha = +,0,-$ and
$\left|i,+\rangle\right. = -\left|\uparrow\uparrow\rangle\right.$,
$\left|i,-\rangle\right. = \left|\downarrow\downarrow\rangle\right.$ and
$\left|i,0\rangle\right. =
(\left|\uparrow\downarrow\rangle\right.+\left|\downarrow\uparrow\rangle\right.)/\sqrt{2}$.
%To preserve the dimension of the Hilbert space, the hard-core constraint
%$\sum_\alpha  t^\dagger_{i\alpha} t_{i\alpha} \leq 1$ needs to be imposed on every site $i$.
%
The Hamiltonian (\ref{H}) can be re-written in triplet
operators.\cite{bondop,kotov,MatsumotoNormandPRL}
The quadratic part reads:
\begin{equation}
\label{H2}
{\cal H}_2 =
\sum_{\vec{q}\alpha}
%\left\{
\left(A_{\vec{q}}-\alpha H\right) t^\dagger_{\vec{q}\alpha}
t_{\vec{q}\alpha} +
\frac{B_{\vec{q}}}{2}\left(t_{\vec{q}\alpha}t_{-\vec{q}\bar{\alpha}}+h.c.\right)
%\right\}
\end{equation}
with $\bar{\alpha} = -\alpha$,
$A_{\vec{q}} = J + B_{\vec{q}}$,
$B_{\vec{q}} = J' (\cos q_x + \cos q_y) + 2 J_z \cos (q_x/2) \cos (q_y/2) \cos q_z$,
where $J_z = J_z^{11} + J_z^{22} - J_z^{12} - J_z^{21}$, and
the field $H$ is in $z$ direction.
%The 3d momentum $\vec k$ runs over the Brillouin zone of the
%bct lattice, spanned by the primitive translations
%$\hat {\bf k}_1=(2\pi,0,-\pi)$, $\hat {\bf k}_2=(0,2\pi,-\pi)$, $\hat {\bf k}_3=(0,0,2\pi)$
%in reciprocal space.
In addition to ${\cal H}_2$ the Hamiltonian contains a quartic in-plane triplet term
${\cal H}_{4\parallel}$;\cite{kotov}
for
$J_{4z} = J_z^{11} + J_z^{22} + J_z^{12} + J_z^{21} \neq 0$ and
$J_{3z}^\pm = J_z^{11} - J_z^{22} \pm (J_z^{12} - J_z^{21}) \neq 0$
quartic and cubic inter-plane interactions, ${\cal H}_{4z}$ and ${\cal H}_{3z}$, arise,
respectively.

\begin{figure}[t]
\epsfxsize=2.8in
\centerline{\epsffile{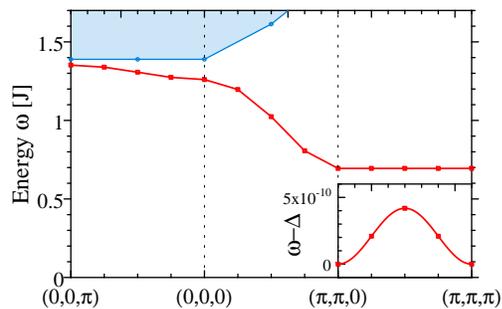}}
%\centerline{\epsffile{./pics/plot1.eps}}
\caption{
(Color online)
Triplon dispersion (solid) and boundary of the two-particle continuum (shaded),
calculated using bond operators,
for $J'/J=0.15$, $-J_z=J_{4z}=0.2J'$, $H=0$.
The inset show an energy zoom into the almost flat vertical dispersion.
The dominant effect of $H$ is a Zeeman shift of $-\alpha H$ of the
triplons.
}
\label{fig:disp1}
\end{figure}

The eigenvalues of ${\cal H}_2$,
$\omega_{\vec{q}\alpha}  = (A_{\vec{q}}^2-B_{\vec{q}}^2)^{1/2} -\alpha H$,
are independent of $q_z$ at in-plane wavevector $\vec{q}_\parallel = (\pi,\pi)$.
In this linearized bond-operator theory, interactions between the order-parameter fluctuations --
represented by triplon quasiparticles -- are ignored.
The most important corrections arise from the hard-core constraint ${\cal H}_U$,
$\sum_\alpha  t^\dagger_{i\alpha} t_{i\alpha} \leq 1$.\cite{kotov}
Following the self-consistent diagrammatic approach of Ref.~\onlinecite{kotov},
we have calculated the renormalized triplon dispersion,
focusing on how interactions lift the degenerate vertical spectrum, Fig.~\ref{fig:disp1}.
Taking into account ${\cal H}_U$ yields a dispersion along $(\pi,\pi,q_z)$ proportional
to $J_z^6 \cos (2 q_z)$, corresponding to the process in Fig.~\ref{fig:dgr1}a.
Further analysis shows that a similar term, but with prefactor $J_z^4$, is
generated to second order in ${\cal H}_{4z}$.
(This cannot be read off from the order-parameter theory ${\cal S}_\phi$ because
of the assumed local form of the quartic term.)
Hence, the bandwidth along $(\pi,\pi,q_z)$ is at most of order $J_z^4$.\cite{tba}
(This result is not changed by ${\cal H}_{3z}$.)

To assess the dynamics of the Ising parameter $\Psi$,
we have studied bound states of two triplons in the singlet channel
by solving a Bethe-Salpeter equation,\cite{kotovbound} with
a four-point vertex as input.
To avoid the necessity for full self-consistency in the two-particle sector,
we have added to ${\cal H}_4$ by hand the biquadratic ${\cal H}_{\rm coll}$ (\ref{hcoll}).
The effective binding force is then given by $(J_{4z}-J_{\rm coll})$;
note that the binding from ${\cal H}_{4z}$ is due to a subtle quantum effect of
singlet formation,
which is not contained in the order-parameter field theory, but is in fact very
common for frustrated spin-1/2 systems.\cite{dimer,tba}

For sufficiently strong positive $(J_{4z}-J_{\rm coll})$ we find a bound state
below the two-particle continuum, whose wavefunction changes sign under 90-degree in-plane
rotations of the internal coordinate, consistent with (\ref{sym}) and Fig.~\ref{fig:schem1}c.
Close to the pressure-driven phase transition
the bound state (with dispersion minimum at total momentum $Q=0$)
is pulled below the {\em single}-particle gap,
due to the weak vertical triplon dispersion.
Clearly, a condensation of this bound state corresponds to the Z$_2$ symmetry breaking
advocated above.
This existence of the bound state depends on microscopic details,
and a comprehensive numerical analysis is difficult due to finite-size effects.
(For the parameter values of Fig.~\ref{fig:disp1}, our calculations indicate no bound
state.)

%%%%%%%%%%%%%%%%%%%%%%%%%%%%%%%%%%%%%%%%%%%%%%%%%%%%%%%%%%%%%%%%%%%%%%%

{\it Pressure vs. field tuning.}
So far, most considerations were for $H=0$.
In finite field, the spin symmetry is reduced to U(1), and the expressions
in (\ref{phi4},\ref{allowed},\ref{forbidden},\ref{phipsi}) are modified accordingly.
However, the symmetry analysis in the paramagnetic phase, leading to
(\ref{allowed},\ref{forbidden}), remains valid.
The most important difference is in the bound-state behavior:
Close to $H_{c1}$ a possible singlet bound state involves at least one high-energy triplet
and will never condense.
Then, scenario (A) always applies, with a 3d QPT which breaks the U(1) $\times$ Z$_2$
symmetry and obeys mean-field exponents.\cite{tba}

%%%%%%%%%%%%%%%%%%%%%%%%%%%%%%%%%%%%%%%%%%%%%%%%%%%%%%%%%%%%%%%%%%%%%%%

{\it Discussion.}
We conclude that dimensional reduction in quantum-critical frustrated
bct magnets does not occur at lowest energies, due to interaction effects.
However, the crossover scale below which 3d behavior is established
(which is given by the vertical dispersion)
is tiny, $E_z \propto J_z^4/J^3$ (instead of $J_z$ as in an unfrustrated system).
Above $E_z$, the shift exponent $\psi$ will take its 2d value.
In the pressure-driven case, depending on microscopic parameters
the ordering transition may be split,
then bond order occurs before magnetic order.

%%%%%%%%%%%%%%%%%%%%%%%%%%%%%%%%%%%%%%%%%%%%%%%%%%%%%%%%%%%%%%%%%%%%%%%

{\it Relevance to \bacu.}
Recent neutron scattering\cite{rueggbacusio} hints on the presence of
{\em multiple} triplon excitations in zero field,
which indicate an enlarged unit cell with inequivalent dimers.
Assuming that the lattice modulation occurs along the $c$ axis,
one arrives at a qualitatively distinct scenario for ``dimensional reduction'':
The condensate established at $H_{c1}$ is strongly inhomogeneous in $z$ direction,
and hence effectively 2d -- note that frustration is {\em not} a required ingredient in
this scenario.
(This conclusion is supported by the fact that the measured vertical triplon
dispersions are tiny for {\em all} in-plane wavevectors,\cite{rueggpriv}
at variance with the results in Fig.~\ref{fig:disp1}.)

%%%%%%%%%%%%%%%%%%%%%%%%%%%%%%%%%%%%%%%%%%%%%%%%%%%%%%%%%%%%%%%%%%%%%%%

We thank I. Fischer, M. Garst, B. Normand, Ch. R\"uegg,
J. Schmalian and especially A. Rosch for discussions.
This work was supported by the DFG SFB 608.

%%%%%%%%%%%%%%%%%%%%%%%%%%%%%%%%%%%%%%%%%%%%%%%%%%%%%%%%%%%%%%%%%%%%%%%

\end{document}